%% file: gcpaper.tex
\documentclass[acmsmall,screen,nonacm]{acmart}

\usepackage{amsmath}
\usepackage{amsthm}
\usepackage{siunitx}

\usepackage{caption}
\usepackage{subcaption}

\usepackage{pgfplots}
\usepackage{pgfmath}
\pgfplotsset{compat=1.18}

\newtheorem{theorem}{Theorem}

\usepackage{url}
\usepackage{listings, lstautogobble}
\usepackage{xspace}
\usepackage{ifthen}

\usepackage{graphicx}
\usepackage{booktabs}

\usepackage{hyperref}
\usepackage[nameinlink]{cleveref}

\newcommand{\gc}{\textsc{Catalpa}\xspace}
\newcommand{\bosque}{\textsc{Bosque}\xspace}

\newboolean{reviewversion}
\setboolean{reviewversion}{false} 

\newcommand{\rvc}[2]{\ifthenelse{\boolean{reviewversion}}{#1}{#2}}
\ifthenelse{\boolean{reviewversion}}{\newcommand{\conclusiontitle}{{Conclusion}}}{\newcommand{\conclusiontitle}{{Onward!}}}

\newcommand{\defaultnursurysize}{\mb{8}}
\newcommand{\defaultpagesize}{\kb{8}}

\newcommand{\pfiftytime}{\us{133}}
\newcommand{\pnintyninetime}{\us{300}}

\newcommand{\eg}{\hbox{\emph{e.g.}}\xspace}
\newcommand{\ie}{\hbox{\emph{i.e.}}\xspace}
\newcommand{\etc}{\hbox{\emph{etc.}}\xspace}
\newcommand{\vs}{\hbox{\emph{vs.}}\xspace}

\newcommand{\wrt}{\hbox{\emph{w.r.t.}}\xspace}

\newcommand{\kb}[1]{\hbox{\qty{#1}{\kilo\byte}}\xspace}
\newcommand{\mb}[1]{\hbox{\qty{#1}{\mega\byte}}\xspace}
\newcommand{\gb}[1]{\hbox{\qty{#1}{\giga\byte}}\xspace}
\newcommand{\secs}[1]{\hbox{\qty{#1}{\second}}\xspace}
\newcommand{\ms}[1]{\hbox{\qty{#1}{\milli\second}}\xspace}
\newcommand{\us}[1]{\hbox{\qty{#1}{\micro\second}}\xspace}

\newcommand{\cf}[1]{\texttt{#1}}

\usepackage{color}
\definecolor{purple}{RGB}{75, 0, 255}
\definecolor{cgreen}{rgb}{0.25,0.5,0.35}

\lstdefinelanguage{bosque}{
keywords={concept, entity, datatype, typedecl, enum, type, provides, field, env, switch, match, abstract, method, if, then, elif, else, function, return, true, false, none, let, var, in, requires, ensures, invariant, validate, recursive, sensitive, using, of, this, pred, fn, ref, examples, for, defer, test, const, override},
keywordstyle=\color{blue}\bfseries,
identifierstyle=\color{black},
alsoother={@},
sensitive=true,
comment=[l]{\%\%},
commentstyle=\color{cgreen}\bfseries
}
 
\setcopyright{acmlicensed}
\copyrightyear{2025}
\acmYear{2025}
\acmDOI{XXXXXXX.XXXXXXX}

\begin{document}

\title{\gc: GC for a Low-Variance Software Stack}

\author{Anthony Arnold}
\email{anthony.arnold@uky.edu}
\affiliation{%
  \institution{University of Kentucky}
  \city{Lexington}
  \state{Kentucky}
  \country{USA}
}

\author{Mark Marron}
\email{mark.marron@uky.edu}
\affiliation{%
  \institution{University of Kentucky}
  \city{Lexington}
  \state{Kentucky}
  \country{USA}
}

\begin{abstract}
The performance of an application/runtime is usually conceptualized as a continuous function where, the lower 
the amount of memory/time used on a given workload, then the better the compiler/runtime is. However, in 
practice, good performance of an application is viewed as more of a binary function -- either the application 
responds in under, say 100 ms, and provides a good user experience~\cite{usability}, or it takes 
a noticeable amount of time, leaving the user waiting and potentially abandoning the task. Thus, performance 
really means how often the application is fast enough to meet user expectations, leading industrial developers to focus on 
the 95$^{\text{th}}$ and 99$^{\text{th}}$ percentile tail-latencies as heavily, or moreso, than average response time.

Our vision is to create a software stack that actively supports these needs via programming language and 
runtime system design. In this paper we present a novel garbage-collector design, the \gc collector, for the \bosque 
programming language and runtime. This allocator is designed to minimize latency \emph{and} tail-latency variability 
while maintaining high-throughput and incurring small memory overheads. To achieve these goals we leverage 
various features of the \bosque language, including immutability and reference-cycle freedom, to construct 
a collector that has \emph{provably} bounded collection pauses, incurs a fixed-constant memory overhead, 
and ensures starvation freedom for the application\rvc{.}{!}
\end{abstract}

\maketitle

\section{Introduction}
\label{sec:intro}
A key-performance-indicator (KPI) for many applications is the $99^{\text{th}}$ (or $95^{\text{th}}$) response percentile latency -- that is, the time it takes for the application to respond 
to a user request $99\%$ of the time. This is a critical metric as these \emph{tail-latency} events are often pain points for users and, once encountered, lead 
to disengagement~\cite{usability}. Unfortunately, these tail-latency events are often intermittent, involve multiple events, and even different subsystems~\cite{tailatscale,toddmeasure}. 
These features combine to make them very difficult to diagnose and resolve.

Some sources of tail-latency are irreducible parts of a distributed (or networked) application, such as connection latency, shared resource contention, or hardware failures. However, 
the latency from these sources is often amplified by runtime and application behavior. For example, a network stall that leads to requests backing up, which leads 
to many objects being promoted into old GC generations, leading to a long GC pause during whole heap collection, causing more requests to back up, and so on. As seen in this 
example, the triggering event for the latency spike is an intermittent network stall but the amplification, along with triage work and resolution, is in the runtime behavior. 

This work focuses on addresssing this challenge via the construction of a programming language and software stack that behave in a fundamentally path independent (or memoryless) 
manner, where regardless of previous operations or even simultaneous executions, the behavior of the systems appears as if each task was executed in isolation. A key aspect of 
creating this type of system is focusing on optimizing for tail-latency variability and avoiding amplifications of pathological behaviors while maintaining high-throughput and 
incurring small memory overheads. This is a radical departure from the current state of the art where modern runtimes are focused on average behaviors and fast path optimization 
but struggle with worst-case behaviors and heuristics. Of particular importance in this area is the design of the memory management and garbage collection system, which are 
often a major source of latency variance, and other performance variability issues in the memory subsystem~\cite{distillingcost,understandcost}.

Leveraging novel aspects of the \bosque programming language~\cite{bosque}, this paper introduces a new garbage collector \gc which is the first language/runtime/gc 
combination capable of satisfying the \emph{no-tradeoff memory subsystem happiness} property (\Cref{thm:happiness}). Recent work~\cite{pathalogical} has theoretically 
validated the conjectures~\cite{understandcost,lxr,urc}, that it is impossible for (mainstream) languages with imperative features to simultaneously ensure bounded 
pause times and starvation-freedom without incurring large performance penalties (thrashing) in other areas. However, \bosque which represents a new viewpoint 
for programming languages, provides a unique opportunity to rethink the design of the memory management system. 

Specifically, this work takes a well known GC design, a copying collector for young objects and a reference counting collector for old objects~\cite{urc,intrc}, and 
simplifies the implementation using the novel aspects of the \bosque programming language to construct a specialized garbage collector with the following novel properties:
\begin{itemize}
  \item \textbf{Bounded Collector Pauses:} The collector only requires the application to pause for a (small) bounded period that is proportional to the size of the nursery.
  \item \textbf{Starvation Freedom:} The collector can never be outrun by the application allocation rate and will always satisfy allocation requests (until true exhaustion).
  \item \textbf{Fixed Work Per Allocation:} The work done by the allocator and GC for each allocation is constant -- regardless of object lifetimes or application behavior.
  \item \textbf{Application Code Independence:} The application code does not pay any cost, \eg read/write barriers, remembered sets, \etc, for the GC implementation.
  \item \textbf{Constant Memory Overhead:} The reserve memory required by the allocator/collector is bounded by a (small) constant overhead.
\end{itemize}

We show that such a system is possible in \Cref{sec:theory} and present a practical design and implementation of the \gc collector in \Cref{sec:algo-overview,sec:implementation}.
Empirically, we validate (\Cref{sec:experimental}) that this design is effective with at $50^{\text{th}}$ percentile GC pause time is less than \pfiftytime with an astonishing 
$99^{\text{th}}$ percentile pause time of under \pnintyninetime and, the reserve memory overhead is proportional to the size of the nursery of \defaultnursurysize. In addition, 
the immutability of values and elimination of read/write barriers leads to an amortized constant cost per allocation and guarantee that each collection can recover either 
all recoverable or at least enough to cover a full cycle of allocation requests (\Cref{sec:algo-overview}). Finally, as shown in \Cref{sec:algo-overview,sec:implementation}, 
the application code does not pay any cost for the GC implementation and the algorithm works with conservative collection~\cite{conservativegc}, enabling the compiler to 
skip root-maps, and easily support pointers into the stack and interior value pointers.

\noindent
In summary the key contributions of this paper are:
\begin{enumerate}
    \item The formalization of the \emph{no-tradeoff memory subsystem happiness} property, along with associated theorems and proofs, as a key objective for modern 
    language/runtime/gc systems (\Cref{sec:theory}). 
    \item A demonstration, via a novel GC construction (\Cref{sec:algo-overview,sec:implementation}), that existing impossibility results~\cite{pathalogical,understandcost} 
    do not apply to languages with the features of \bosque.
    \item An experimental evaluation of the collector showing that, in addition to its theoretical guarantees, the combined language/runtime system achieves (very) low and 
    predictable pause times along with low memory overheads in practice (\Cref{sec:experimental}).
\end{enumerate}

\section{\bosque Background}
\label{sec:background}
\input{background}

\section{GC Algorithm Overview}
\label{sec:algo-overview}
\input{algoverview}

\section{\gc Implementation}
\label{sec:implementation}
\input{implementation}

\section{Theoretical Analysis}
\label{sec:theory}
\input{theory}

\section{Experimental Evaluation}
\label{sec:experimental}
\input{experiemental}

\section{Related Work}
\label{sec:related}
\input{related}

\section{\conclusiontitle}
This paper presents a novel garbage collector design for the new \bosque programming language and runtime. A key design objective in this project 
generally, and this collector specifically, is to create a software stack that provides predictable and low-latency performance along with a 
very light memory footprint and small tail latencies. The \gc collector presented in this this work is a key component in this systems and, 
represents the first language/runtime/gc combination capable of satisfying the \emph{no-tradeoff memory subsystem happiness} property (\Cref{thm:happiness}).
The experimental results provide strong preliminary evidence that the theoretical properties of the collector are borne out in practice. 
As a result, we believe that this work represents a significant development in design of memory management systems for modern applications 
and opens up a new area of research in the design of runtime and GC systems focused on the (reliability and stability) requirements of modern software 
systems.

\section*{Data Availability}
All code, data, and benchmarks used in this study are publicly available and open source (MIT) licensed at \url{https://github.com/BosqueLanguage/BosqueCore}. 

\bibliographystyle{ACM-Reference-Format}
\bibliography{bibfile}

\end{document}

%% file: background.tex

The goal of the \bosque project is to create a programming system that is optimized for reasoning -- by humans, symbolic analysis tooling, 
and AI agents (Large Language Models in particular)~\cite{bosque}. The approach taken by \bosque is to identify and remove features or concepts 
that complicate various forms of reasoning and that are frequent causes of software faults, increase the effort required for a developer (or AI agent) 
to reason about and implement functionality in an application, or complicate automatically reasoning about a program. 
Although the initial motivation of this work was focused on software assurance and quality, these same simplifications also provide strong 
guarantees about how memory can be allocated, organized, and used at runtime as well\rvc{.}{!} 

At the core of \bosque is a let-based functional language with a nominal type system for declaring datatypes. A sample \bosque program for 
computing the largest low-high temperature range in a list is shown in \Cref{fig:temp-range}.

\begin{figure}[ht]
\begin{lstlisting}[language=bosque]
type Fahrenheit = Int;

entity TempRange {
  field low: Fahrenheit;
  field high: Fahrenheit;

  invariant $low <= $high;
}

function maxTempRange(temps: List<TempRange>): TempRange {
  return temps.maxElement(pred(t1, t2) => {
    return t1.high - t1.low < t2.high - t2.low
  });
}

maxTempRange(List<TempRange>{
    TempRange{32<Fahrenheit>, 50<Fahrenheit>},
    TempRange{40<Fahrenheit>, 60<Fahrenheit>},
    TempRange{20<Fahrenheit>, 30<Fahrenheit>}
});

%% Result is TempRange{40<Fahrenheit>, 60<Fahrenheit>}
\end{lstlisting}
\caption{Max Temperature Range in the \bosque Programming Language.}
\label{fig:temp-range}
\end{figure}

The first declaration in the code in \Cref{fig:temp-range} is a type declaration for a new type \cf{Fahrenheit} that is an alias for the \cf{Int} type. 
This allows the creation of a new type that is distinct from \cf{Int} but has the same (efficient) underlying representation. Next is a \cf{entity} 
declaration of a composite datatype \cf{TempRange} that has two fields: \cf{low} and \cf{high}, both of type \cf{Fahrenheit}. The \cf{invariant} declaration 
ensures that the \cf{low} field is always less than or equal to the \cf{high} field whenever a \cf{TempRange} value is created.

The function \cf{maxTempRange} takes a list of \cf{TempRange} values and returns the one with the largest difference between the \cf{high} and \cf{low} 
temperature fields. The code uses a higher-order function \cf{maxElement} that takes a predicate function to compare two \cf{TempRange} values. The last 
expression in the code is a call to \cf{maxTempRange} with a literal list of three \cf{TempRange} values. The result of this call is a \cf{TempRange} 
with the value \cf{TempRange\{40<Fahrenheit>, 60<Fahrenheit>\}}.

There are a number of aspects of this example that are interesting from a memory management perspective.

\paragraph{\bf Immutability:}
The \cf{entity} declaration in \bosque creates a new composite datatype and these are always immutable. A key implication of this fact is that once 
an entity value is created then fields (and thus pointer targets) will never change.

\paragraph{\bf Referential Transparency:}
The semantics of \bosque ensure that reference identity of values is never observable -- either directly via equality tests or indirectly via mutation. 
Thus, the allocator has wide latitude in value representation and placement. Values can be stack, heap, or inline allocated without concern and object 
relocation is guaranteed to be semantically safe.

\paragraph{\bf Cycle Freedom:}
In combination with the immutability of entities, and careful definition of constructor semantics, the \bosque language 
ensures that all data structures are acyclic. This invariant allows us to utilize reference-counting techniques without concern for backup cycle-collection 
or other special case logic.

\paragraph{\bf Non-Escaping Lambdas:}
Although \bosque supports first-class functions and higher-order functions, and applications use them heavily, the language semantics require them to be in 
direct argument position as literals or passed parameters. As a result a lambda function cannot escape from the scope in which it is defined and the compiler 
can fully monomorphize higher-order code.

\paragraph{\bf Ropes and RRB-Vectors:}
Lists and Strings in \bosque are implemented via efficient tree-structures~\cite{persistentrrb}. This design allows for efficient List/String processing, 
including appends and inserts, and also has the benefit that even large strings (or lists) are implemented as a tree of fixed-size chunks. This, along with
the closed-world compilation model, implies that all allocation sizes can be computed at compile time and are all small values.

%% file: algoverview.tex
The overall design of the garbage collector is based on a hybrid-generational approach~\cite{urc}, with a copying nursery for young objects and promotion 
to a reference counted old-space. For the stack we use a conservative scan~\cite{conservativegc} while objects are handled precisely. The top-level algorithm 
for the allocator and collect phase of the implementation are shown in \Cref{fig:collectoverview}.

\begin{figure}[ht]
\begin{lstlisting}[language=c++]
template <size_t K>
class Allocator 
{
    FreeEntry<K>* freelist;

    ...

    void* alloc(Type* t) 
    {
        entry = this->freelist;
        if(this->freelist == nullptr) {
            return allocSlow(type); //collector is run from here if/as needed
        }

        this->freelist = this->freelist->next;
        return INIT_META(entry, type);
    }
};

...

void collect()
{
    markRoots();
    markHeap();
    processMarkedYoung();

    computeDeadRootsForDecrement();
    processDecrements();
}
\end{lstlisting}
\caption{Top-level allocation (size-segmented) and collection algorithms. The \cf{Allocator} class is templated on object sizes \cf{K} and maintains a free-list of available
locations, per memory page, to allocate from in the \cf{freelist} field. The \cf{collect} function shows the high-level steps of the collection algorithm which is run when 
the nursery is full (default of \defaultnursurysize).}
\label{fig:collectoverview}
\end{figure}

The code in \Cref{fig:collectoverview} shows the key features of the allocator code. The closed world semantics of \bosque enable us to pre-compute the sizes of every 
allocation needed in an application and statically create dedicated (thread-local) allocators for each size class. To perform an allocation each \cf{Allocator} maintains a 
page of memory to allocate from and this page is organized using a single free-list layout of available locations. When an allocation is requested, the allocator 
has a fast path of taking the head of the \cf{freelist} and advancing the \cf{next} pointer. If the \cf{freelist} is empty (\cf{nullptr}) then the allocator 
calls a slow path to allocate a new page of memory, or run a collection cycle, and initialize the freelist for that page.
This design provides good baseline allocator behaviors in terms of inlineable fast-path allocation performance and, the use of size-segmented and per-page 
free-lists, provides good spatial locality for object allocations~\cite{immix,mimalloc}.

In our collector design heap values can be in one of two logical regions, the nursery or the reference-counted old space. This is shown in \Cref{fig:memlogicallayout}. 
In this design the roots can point to values in the nursery, the RC-heap, or other stack allocated values\footnote{As described in \Cref{sec:implementation} roots may 
also point to the interior of stack values or heap allocated objects.}. However, in contrast to most generational collectors, there may be pointers from the nursery to the 
RC-heap \emph{but not} pointers from the RC-heap to the nursery. This hierarchical relation allows for simplified garbage collection, as 
objects in the nursery can be collected without concern for references from older objects which cannot exist\rvc{.}{!}

The code in \Cref{fig:collectoverview} shows the high-level overview of the collection algorithm. The first step is to mark all root references, which includes 
conservatively scanning the stack, global variables, and registers for pointers to heap objects. This code uses standard conservative scanning methods~\cite{conservativegc,chakracore}. 
The next step is to mark all reachable objects in the nursery, the \cf{markHeap} call, which starts from the root set, traversing the object graph, and marking all 
reachable \emph{young} objects (see \Cref{sec:implementation} for details).

\begin{figure}[t]
\centering
\begin{subfigure}[t]{0.45\columnwidth}%
    \includegraphics[width=\textwidth]{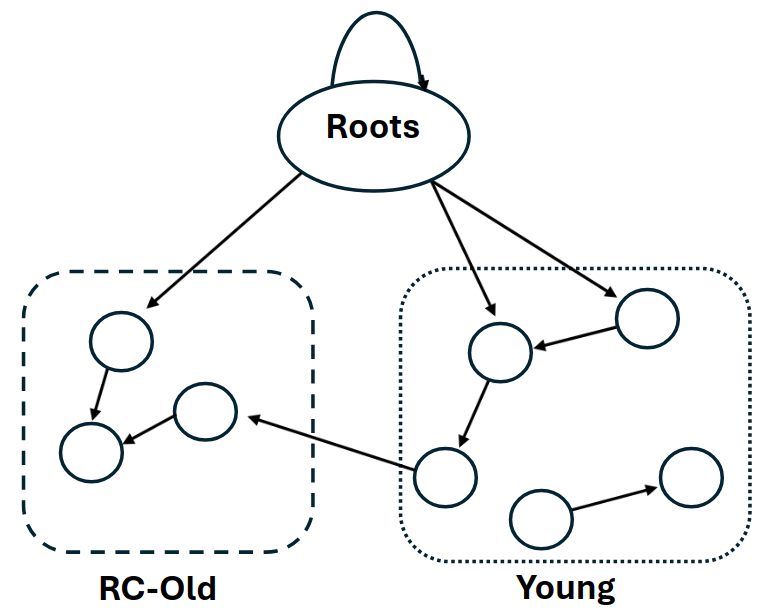}%
    \caption{State of memory at the start of a collection cycle. Note root references may point to stack locations, nursery, or RC-heap objects \emph{but} pointers from the 
    old RC-heap to the nursery are \emph{not} possible.}%
    \label{fig:memlogicallayout}
\end{subfigure}
\hspace{7mm}
\begin{subfigure}[t]{0.45\columnwidth}%
    \includegraphics[width=\textwidth]{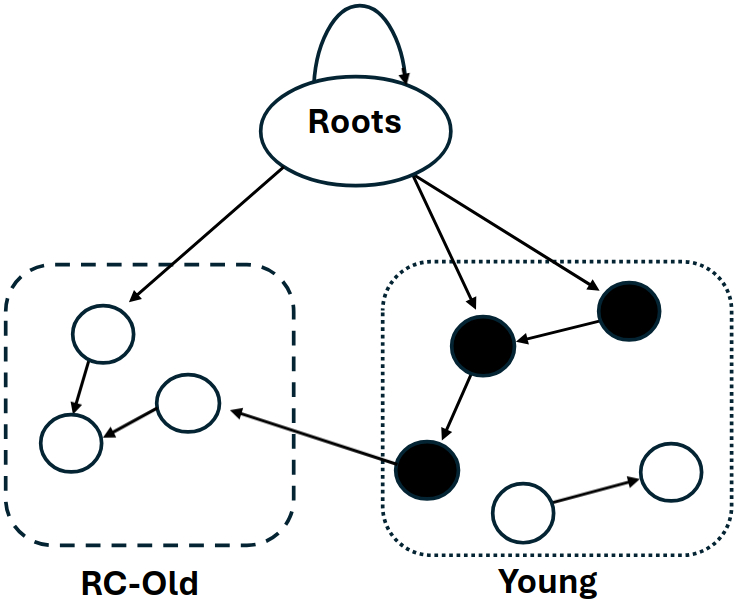}%
    \caption{State of memory after marking young objects. The marked objects are shown colored in black.}%
    \label{fig:aftermarkyoung}
\end{subfigure}
\vspace{7mm}
\\
\begin{subfigure}[t]{0.45\columnwidth}%
    \includegraphics[width=\textwidth]{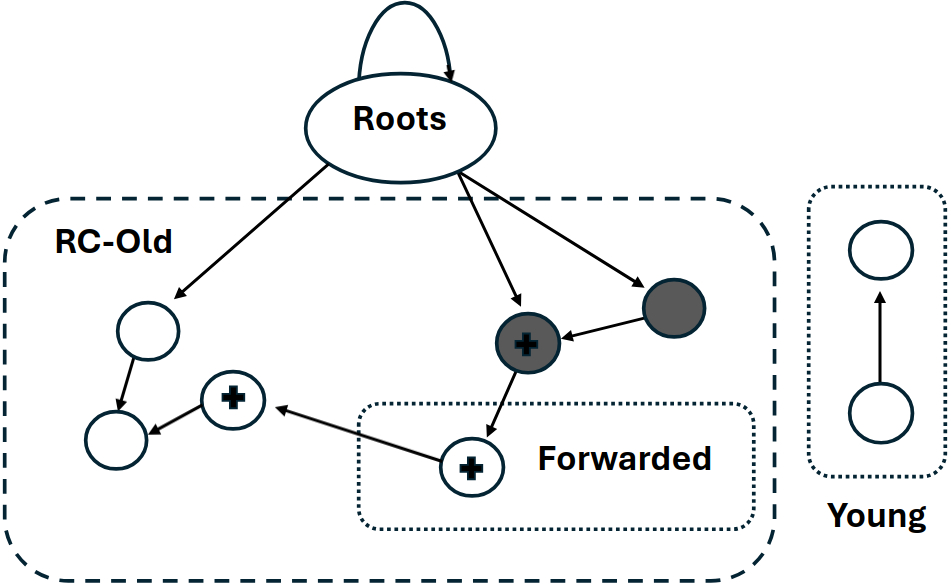}%
    \caption{State of memory after processing marked objects. The young objects are evacuated and compacted to pages in the old-space if possible (shown in the forwarded region) or promoted to RC representations in 
    place if they must be conservatively handled (\ie there is a root reference to them). Objects with reference count increments are shown with a $\oplus$ symbol.}%
    \label{fig:aftermarkprocess}
\end{subfigure}
\hspace{7mm}
\begin{subfigure}[t]{0.45\columnwidth}%
    \includegraphics[width=\textwidth]{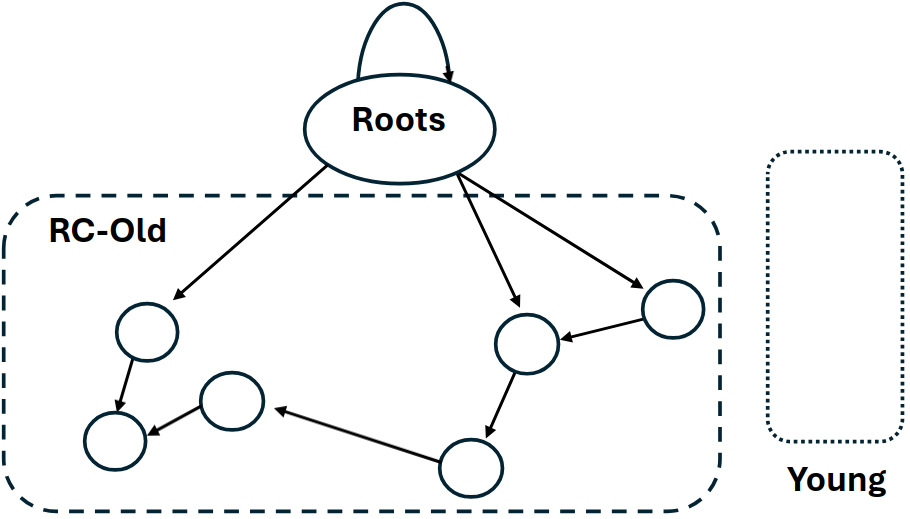}%
    \caption{State of memory after decrements and reset nursery. All of the unreachable objects from the young space have been reclaimed (via freelist management) and RC object with counts of 
    $0$ are collected.}%
    \label{fig:afterdecs}
\end{subfigure}
\caption{Evolution of the (logical) memory layout during a collection cycle -- from initial state, to marking young objects, evacuating young objects, adjusting reference counts,
and finally resetting the nursery.}%
\label{fig:memlogicallayout}
\end{figure}

The state of (logical) memory model after these marking steps is shown in \Cref{fig:aftermarkyoung}. In this figure the nursery is shown with a set of marked objects 
(shaded in black) that have been identified as reachable from the root set. Note that objects in the RC-old space are not marked and are never visited during the marking phase.

The next step is to process the marked young objects. For each marked object there are two possibilities, the first is that the object is referenced by a 
conservative root location, and the second is that the object is referenced only by other young objects. In the first case we cannot relocate the object, 
as the root location is conservative and cannot be updated. Thus, these objects are converted in place to a reference counted representation, shown as 
now black in \Cref{fig:aftermarkprocess}. In the second case we can evacuate the object values to a compacting page, performing pointer forwarding as necessary, 
and converting them to reference counted representations (the objects in the dashed box). As each object is promoted to the RC-old space all fields are 
(precisely) scanned and reference counts for children are incremented (shown with black plus in \Cref{fig:aftermarkprocess}).

Next the algorithm sweeps the now mostly (or entirely) evacuated nursery and rebuilds the free-lists for the next round of allocation -- allowing the collector to 
re-use both completely empty and semi filled pages. As RC objects are not moved the ability to re-use partially filled pages is critical to avoid fragmentation. 
The sweep reclaims the unmarked dead objects and the space cleared by the evacuation. In our example, there is only one object 
that remains on the page, the root referenced object, while all other slots are evacuated or reclaimed (\Cref{sec:implementation}).

The final step in the collection algorithm is to process deferred reference count decrements and root reference status. Our implementation compares the root set 
from the previous collection with the root set identified in the current collection to determine which root references are no longer live. For each of these references 
the collector checks if the reference count has dropped to zero, in which case the object can be reclaimed. This reclamation is then a standard release 
walk of the object graph, decrementing reference counts and reclaiming objects as necessary. The state of the memory after this step is shown in \Cref{fig:afterdecs}.
As each object is reclaimed, its memory is returned to the free list, in the appropriate page for later use.

The most notable feature of this algorithm is not what it has but what doesn't. Despite the use of a generational collector and reference-counting, there are no 
remembered sets, no write barriers, no backup cycle-collector, and no need for runtime support in the application code. As a result of these features, the cost of a 
collection cycle is independent of any application code behavior and can always be performed in a bounded time that is proportional to the size of the 
nursery -- and the additional memory overhead is also a constant factor of the nursery size\rvc{.}{!}

%% file: implementation.tex

The logical model as described in \Cref{sec:algo-overview} provides a clean separation between
the nursery and RC-old space. However, a practical implementation must map these logical regions 
onto physical pages in memory in a manner that is efficient for both allocation and collection. 
This section details that mapping, the management of physical pages, and key implementation details 
of the processing algorithms.

\subsection{Memory Organization}
The \gc memory management system is based on a set of thread-local page pools which feed a set of 
size-segregated allocators as shown in \Cref{fig:allocmodel}. As \bosque provides a closed-world 
compilation model, and all lists/strings are implemented as trees/ropes~\cite{chakracore,persistentrrb}, 
the compiler can pre-compute all allocation sizes. Thus, we do not need to handle large, or variable 
sized, objects.  

Each allocator is responsible for a given size class and, in the active allocation page (\cf{allocPage} in 
\Cref{fig:allocmodel}), the allocator maintains a null terminated free-list of available slots for allocation.
The allocator also tracks pages used for evacuation during collection (\cf{evacPage} in \Cref{fig:allocmodel}) 
and a set of partially filled pages organized by utilization.

This free-list organization is required as the collector performs a conservative stack scan and the reference 
counted old-space objects are not movable. Since the collector may need to repurpose partially filled pages as
allocation pages, we employ a free-list-based bump-pointer design where each entry corresponds to a fixed-size
slot determined by its allocator. This ensures every free slot remains available for allocation, providing strong
defense against fragmentation. The current implementation stores allocation metadata inline in the object header 
although out-of-band metadata may be desirable in a multi-threaded implementation to minimize false sharing.

\subsection{Allocation}
Value allocation is done into a nursery using a thread-local (per page) free-list pointer (\Cref{fig:collectoverview}) 
that is segregated based on object size. The head of the active allocation free-list is also stored 
directly in the thread-local static section. This combination of thread-locality, size-segregation, 
direct free-list access, compile time known sizes allows us to emit an efficient fast path inline
at each allocation site.

\begin{figure}[ht]
\centering
\includegraphics[width=0.7\columnwidth]{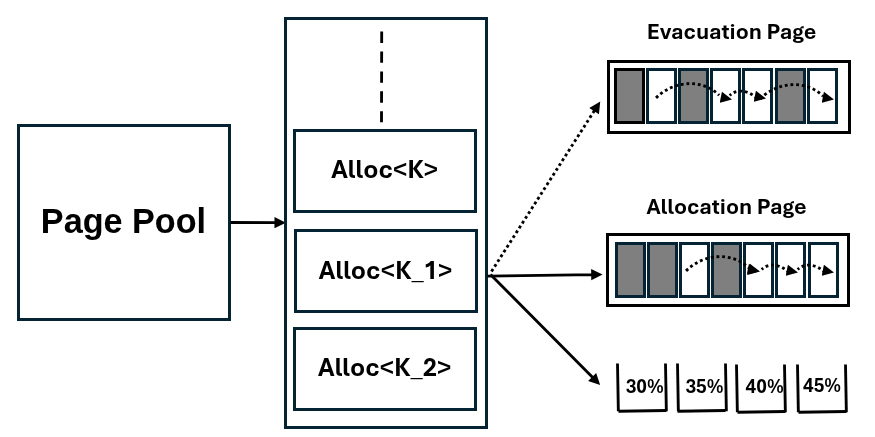}
\caption{\gc Memory organization of pages, size-segregated allocators, and per-page free-lists. Each allocator manages a set of pages 
for allocation/evacuation, and partially filled pages for the RC-old space.}
\label{fig:allocmodel}
\end{figure}

This fast path quickly checks if the free-list is non-empty, and if so takes the head of the free-list,
advances the free-list pointer, and initializes the object metadata. If the free-list is empty then the 
allocator calls a slow-path function that will either scavenge a new page of memory from the page pool,
or trigger a collection cycle to reclaim memory. As shown in \Cref{fig:allocmodel} the allocator maintains a 
list of partially filled pages in the page pool which can be selected as the new allocation page, falling 
back to requesting a new page from the OS if none are available.

In the \gc design we select from the set of partially filled pages the one with the lowest utilization for 
allocation and evacuation. This design ensures that pages are filled up as much as possible before new pages are
requested from the OS, improving spatial locality and reducing overall memory usage. As reference-counted objects 
are naturally reclaimed over time the allocator will slowly compact pages and free up fully empty pages for 
return to the OS.

\subsection{Young Collection}
A collection trigger is included on the allocator slow-path and is controlled by a fixed nursery threshold, 
by default \defaultnursurysize. The stack scan is performed conservatively, in contrast to the precise object scan, and 
allows for stack allocated objects as well as interior pointers into stack/heap allocated objects. This 
design allows the compiler to avoid root-maps, allows aggressive value conversion (which is possible as 
\bosque is fully referentially transparent), and allows for easy inter-operation with C/C++ code.

Once the root set is computed the collector runs a tracing collection of the nursery. As the stack scan is 
conservative we cannot automatically evacuate all young objects, as we cannot safely relocate possible 
stack pointers. Thus, we either promote each young object in place, setting metadata to \emph{old} with a 
initializing the reference count, or evacuate the object including setting the metadata and setting up a forwarding pointer. 
As we move objects we leverage the invariants that there are \emph{no old$\rightarrow$young} pointers and that
the object graph is acyclic, which enables us to process young objects in reverse topological order and 
ensure all pointers are updated correctly. In the case of a \emph{young$\rightarrow$old} pointer we simply 
perform a reference count increment operation as the young object is currently being promoted to the RC 
old-space.

\subsection{Reference Count Management}
To start the RC phase, \gc begins by first comparing the root set collected in the previous collection 
with the set observed in the current collection. These sets are ordered based on address and a linear 
two-pointer walk is performed to determine whether a root object is present in the old set, but not the 
current set. If an object reference count is zero, and it is not referenced by a root, it is then inserted 
on a worklist for pending RC decrement operations and reclamation.

We process this worklist of now dead objects by decrementing the reference count of each child, enqueuing any dead 
child objects onto the same worklist, and then releasing the object itself. 
This worklist design allows us to control the amount of work done in a given collection, 
stopping if there is a risk of a large deletion cascade. Currently, the \gc collector will stop processing 
after releasing $1.5 \times N$ objects, where $N$ is the approximate number 
of objects distributed out of pages in the previous round of allocations. This partial processing with 
bounded work per cycle (or heartbeat~\cite{mimalloc}) ensures that the collector always blocks 
for a bounded period, while ensuring that reclaimed memory is proportional to the 
allocation rate, and ensures we never starve the application.

As the overall \gc collector is non-moving, and RC old-space objects from any page can die in any collection, 
then as time progresses pages will slowly drop in their utilization. As the allocator distributes objects it 
will require new pages to draw from. Thus, as shown in \Cref{fig:allocmodel}, this set of partially filled pages 
is where the allocator can select a new allocation page from. For best performance we want to select a page 
that has a reasonable number of free slots, amortizing the cost of the slow-path work over as many allocations 
as possible. 

To keep an accurate and up-to-date view of page utilization we update this information with defered 
live count updates during the RC processing. We organize our pages into utilization bins ($5\%$ increments)
to prevent frequent moves between categories. Each bin is maintained as an approximate set based on page 
utilization. When a page's computed utilization crosses a predetermined threshold it is moved into the 
appropriate bin.

A key observation in this algorithm is that there is no cycle-collector needed as \bosque ensures that 
reference cycles cannot be created. This ensures that the RC processing is always over DAG and all decrements 
will eventually terminate with all unreachable object reference counts going to $0$ and being reclaimed.

%% file: theory.tex

This section provides a theoretical analysis of the properties and guarantees of the overall memory management system.

\begin{theorem}[Fixed Work Per Allocation]\label{thm:fixedwork}
    The work done by the garbage collector (GC) for each allocation is fixed and bounded by the function 
    $O(\text{Field}_{\text{ct}} * (\text{Cost}(\text{Mark} + \text{Fwd} + \text{Inc} + \text{Dec}) + \text{Cost}(\text{Alloc} + \text{Copy} + \text{Release})))$, 
    and does not depend on the lifetime or behavior of the application -- asymptotically this is $O(1)$.
\end{theorem}

\Cref{thm:fixedwork} states that the work done for each allocation is constant and is independent of the actions of the application 
code. The \text{Alloc} term represents the initial cost to allocate the object in the nursery and, is a simple pointer 
operation. After allocation the next lifecycle phase is collection and (possible) promotion to the old reference-counted generation. During a collection 
each allocated object (if still alive) will be visited during the marking and evacuation. During the marking phase we visit each field of the object, 
for $\text{Field}_{\text{ct}} * \text{Cost}(\text{Mark})$ work and during the evacuation we perform forwarding and (as needed) increments to 
field values pointers for $\text{Field}_{\text{ct}} * \text{Cost}(\text{Fwd} + \text{Inc}) + \text{Cost}(\text{Copy})$ work.

Once in the old-space cycle-freedom eliminates the need for a cycle-collector and immutability eliminates the need for a remembered set and the possibility 
of re-processing. These two features ensure that old-value objects will never be re-visited during liveness/reachability processing.

While the object is live in the old-space there may be increments and decrements of its reference count. The \text{Inc} operations are already accounted for 
as part of the promotion cost of predecessor objects. Decrements will only occur when a predecessor object is reclaimed. As with the \text{Inc} operations, the \text{Dec} operations 
can be accounted for as part of the predecessor reclaim cost. This cost is then $\text{Cost}(\text{Release}) + \text{Field}_{\text{ct}} * \text{Cost}( \text{Dec})$. 

\begin{theorem}[Bounded Collector Pauses]\label{thm:boundedpauses}
    The garbage collector (GC) pause times are bounded by a constant factor $K$, determined by the size of the nursery,
    and are independent of application behavior or allocation rate.
\end{theorem}

\Cref{thm:boundedpauses} states that the garbage collector (GC) pause times are bounded by a constant factor $K$ determined by the size of the nursery. 
As the GC runs in two phases -- processing, and possibly promoting, young objects in the nursery followed by checking root sets and updating reference counts 
-- the cost of a collection is proportional to the work to process the nursery ($K_{\text{nursery}}$) and the cost to process roots and perform the 
reference operations ($K_{\text{old}}$).

The cost of the nursery component $K_{\text{nursery}}$ is the cost to mark and evacuate live objects + the cost to sweep the nursery and rebuild free-lists. 
As the number of references/objects to process is fixed by the size of the stack/statics, as all objects are immutable and there is no remembered set, the 
number of objects marked and/or evacuated is bounded by the size of the nursery. The cost to sweep the nursery and rebuild free-lists is also, by construction, 
proportional to the size of the nursery. 

The cost for performing root set reference operations, $K_{\text{old}}$, is a function of the number objects that had a root reference in the previous collection 
but do \emph{not} have a root reference in the current stack. This is bounded by the size of the application stack. A naive decrement/release walk of 
these objects could, in the worst case, touch all objects in the heap. However, as described in \Cref{sec:implementation}, we use a classic control system to perform 
this work (possibly) over multiple collection cycles while ensuring that the overall cost remains bounded per collection cycle and, 
also, that we monotonically decrease the number of pending decrements.

\begin{theorem}[Effective Collections]\label{thm:starvation}
    After a collection completes it has either reclaimed all unreachable objects or at least $1.X\times$ the size of the nursery -- 
    as a result the application allocation rate can never outrun the collector.
\end{theorem}

\Cref{thm:starvation} states that our collection strategy is always effective in reclaiming memory and that the application will never starve for memory.
As a corollary of the proof for \Cref{thm:boundedpauses} we have that either all reclaimable objects have been identified and recycled at the end of the 
collection \emph{or} we have recycled at least the amount of memory allocated in the nursery plus a fraction given by $X$. This ensures that the application can never outrun the collector.

\begin{theorem}[Fixed Memory Overhead \wrt Application Memory Usage]\label{thm:overhead}
    The memory overhead of the system \wrt to the live memory usage is given by a constant factor $K$ determined by and 
    proportional to the size of the nursery.
\end{theorem}

\Cref{thm:overhead} states that the memory overhead of the system \wrt to the live memory usage is given by a constant factor $K$ determined by and proportional to 
the size of the nursery. As described in \Cref{sec:algo-overview} the collector uses a nursery for the young space and all promoted objects are handled via a 
reference counting mechanism. Thus, the size of the old space is proportional to the size of dynamically live application objects. As the nursery is a fixed size, 
and the book-keeping data structures described in \Cref{sec:implementation} are also proportional to the size of the nursery, the overall overhead is a 
constant factor \wrt the live application memory usage.

\begin{theorem}[Memory Subsystem Happiness]\label{thm:happiness}
    The allocation rate of the application can never outrun the garbage collector (GC) and the collector only touches objects on the \emph{fringe} 
    of the old reference-counted space -- thus starvation is eliminated, pause times are bounded, collection is always effective, and GC driven cache/page eviction 
    is minimal.
\end{theorem}

As a result of the above theorems we have \Cref{thm:happiness}. This summarizes the properties of the allocator/collector as a whole and the 
unique \emph{no-tradeoff} nature of the result. Additionally, as the collector never touches objects in the old reference-counted space unless an object 
that is being promoted from the nursery references it \underline{or} an object in the reference-counted space that references it is being reclaimed. Thus, 
these \emph{fringe} objects are the boundary for memory touched by the collector in the old space -- and any interior objects will never be accessed by 
the collector. This ensures that the collector does not thrash the memory subsystem. 

\Cref{thm:happiness} along with the analysis in~\cite{pathalogical} demonstrate that \bosque (and the \gc runtime) are unique in satisfying the \emph{no-tradeoff 
memory subsystem happiness} property. Further, as proved in~\cite{pathalogical} it is theoretically impossible for any (mainstream) language with imperative features 
to achieve this level of performance without significant trade-offs in other areas!

%% file: experiemental.tex

As described in \Cref{sec:intro} our focus in this work is on creating a software stack with highly-predictable performance characteristics. 
Thus, our evaluation in this section is split into three parts. The first section provides a general analysis of application performance and GC 
specific components. The second is a brief analysis of the sensitivity of the \gc collector to its configuration parameters. Finally, the 
third section provides a detailed set of experiments to evaluate end-to-end latency of the workloads as well the path-independence behaviors 
of the overall system.

As \bosque is a new language, there are limited existing applications to use as benchmarks. Thus, we focus on a commonly used sample applications 
that have been re-implemented in \bosque. The first is a \bosque implementations of the n-body simulations programs from the Computer 
Language Benchmarks Game~\cite{bencharkshoot}. The next is a raytracing program published on Microsoft's MSDN blog~\cite{msdnraytrace}. The db program is 
a \bosque implementation of the DB benchmark from SpecJVM 98~\cite{specjvm98}. The final benchmark is the optimizer pass of the \bosque compiler (written in \bosque) 
which is the most complex of the benchmarks and largest \bosque program currently in existence.

The \gc collector is implemented in C++, $2.4$kloc at present, and all evaluations use the standard release build configuration.
All experiments were run on a system with an AMD Ryzen 9 9950X and 64GB of memory. The system is otherwise unloaded to minimize 
the impact of other workloads on the performance measurements. All runs use a default nursery size of \defaultnursurysize and a default page 
size of \defaultpagesize.

\begin{table*}[ht]
\centering
\begin{tabular}{cccccc}
\toprule
Benchmark & Code Size & Types & AllocCount & AllocMemory (\gb{}) & Max Live Heap (\kb{}) \\
\midrule
n-body    & 193 & 68 & 1,248,474,177 & 69.5 & 5.1 \\
raytracer & 273 & 34 & 822,135,153 & 34.4 & 2.8 \\
db        & 304 & 71 & 1,970,703,992 & 92.3  & 46.9 \\
compiler  & 5120 & 684 & 1,538,395,937 & 140.9 & 8,519 \\
\bottomrule
\end{tabular}
\caption{Static and dynamic statistics for the evaluation applications. The Code Size column is lines of \bosque source code and Types is the number of distinct \bosque types in the 
program. The next three columns are dynamic memory statistics -- AllocCount is the total number of allocations, AllocMemory is the total bytes allocated, and the Max Live Heap 
column is the max live heap observed during execution.}
\label{tab:macrostats}
\end{table*}

\Cref{tab:macrostats} shows a set of static and dynamic statistics for each of the benchmarks. The code size column indicates the number of 
lines of \bosque source code for the benchmark while the types column indicates the number of distinct \bosque types created in the program. 
The next three columns provide dynamic memory statistics for the execution of the benchmark -- specifically the total number of allocations 
made in the execution, the total number of bytes allocated, and the max live heap observed during the execution. 

The statistics in \Cref{tab:macrostats} show that the benchmarks allocate heavily during their execution, primarily as a result of the direct 
implementation of immutable data structures in \bosque. In these structures each modification performs a logarithmic number of allocations as a 
partial new list is created and mapping (or filtering) over a list allocates a fresh structure. This results in a high allocation rate of 
small objects making it critical that the GC can keep up with the workload without stalling. 

The Max Live Heap values \Cref{tab:macrostats} are computed based on the meaningful data values in the application code
\emph{excluding} code pages, memory allocator/collector metadata, or other runtime data structures. These, relatively small live heaps, 
are similar to the memory footprints seen in many types of microservice workloads, such as Node.js~\cite{node} or serverless 
applications~\cite{serverlessdef}, which have relatively small memory footprints, and keeping memory overheads and resource provisioning 
work low is critical to efficient operation.

\subsection{GC Performance}
\label{sec:macro-performance}
The first evaluation is a throughput comparison between \gc and an $\epsilon$-gc collector which allocates continuously from a bump 
buffer without any collections. We measure the total wall-clock time taken by the application and pause times for the collector. These results 
are shown in \Cref{tab:macrooverall}. The first column is the benchmark name, the second and third columns are the total wall-clock time for 
the application to run with the \gc collector and the $\epsilon$-gc collector respectively. 

\begin{table*}[ht]
\centering
\begin{tabular}{cccccc}
\toprule
Benchmark & \multicolumn{2}{c}{Application Time (s)} & \multicolumn{3}{c}{\gc GC Pause (\us{})} \\
& \gc & $\epsilon$-gc & 50\% & 95\% & 99\% \\
\midrule
n-body & 1.14 & 1.72 & 137 & 172 & 186 \\
raytracer & 1.03 & 1.18 & 151 & 198 & 214 \\
db & 1.16 & 1.66 & 170 & 216 & 231 \\
compiler & 1.13 & 1.64 & 166 & 213 & 258 \\
\bottomrule
\end{tabular}
\caption{The first two columns compare the wall-clock time for executing the application under the \gc \vs $\epsilon$-gc collectors. The following three columns cover the 
\gc collector pause time statistics at three standard measures, $50^{\text{th}}$ percentile, $95^{\text{th}}$ percentile and $99^{\text{th}}$ percentile.}
\label{tab:macrooverall}
\end{table*}

The results in \Cref{tab:macrooverall} show that in all cases the \gc collector is actually faster than the $\epsilon$-gc collector by on average $25\%$. 
Our analysis indicates that this is due to the improved locality of the memory access patterns after copy-compaction out of the nursery which more than offsets
the additional work required by the collector. The average pause times for the collector are quite low, with a $50^{\text{th}}$ percentile pause time of 
\us{137}-\us{170} across the benchmarks and a $99^{\text{th}}$ percentile pause time of \us{258} on any benchmark! 

We also observe that the pause times are quite consistent across the benchmarks. Thus, as expected from the theoretical analysis in \Cref{sec:theory}, we 
see that the collector performance is largely invariant of the application workload and primarily a function on the nursery size.

\begin{table*}[ht]
\centering
\begin{tabular}{ccccc}
\toprule
Benchmark & Collections & Survival Rate & GC \%Time & Heap Size (\mb{}) \\
\midrule
n-body & 758 & 0.03\% & 10.5\% & 8.6 \\
raytracer & 245 & 0.007\% & 3.7\% & 8.6 \\
db & 553 & 0.31\% & 12.4\% & 8.7 \\
compiler & 615 & 0.36\% & 9.5\% & 16.1 \\
\bottomrule
\end{tabular}
\caption{The first two columns show the total number of collections performed by the \gc collector during the benchmark run and 
the average survival rate of the nursery (at \defaultnursurysize). The next column shows the percentage of total application time
spent in GC. The final column shows the max memory used by the application, runtime, \emph{and} collector as measured by total page usage 
from the OS.}
\label{tab:macrooverehead}
\end{table*}

Critically, the temporal behavior seen in \Cref{tab:macrooverall} is \emph{not} achieved at the expense of memory 
overheads~\cite{understandcost}. As shown in the last column of \Cref{tab:macrooverehead}, the maximum \emph{heap size}
used by the application during the execution of the benchmark, measured as the size of all committed memory pages used in the computation, 
is under \mb{17} for every application -- or only slightly more than the nursery size, \defaultnursurysize, plus the live heap size. 
This indicates that in practice the memory overheads of the \gc collector are quite low and closely track the theoretical expectations.

As expected from the rapid rate of allocation seen in \Cref{tab:macrostats}, the number of collections performed by the \gc collector 
is high (column 2 of \Cref{tab:macrooverehead}) and the local/temporary nature of the allocations leads to low survival rates
(column 3 of \Cref{tab:macrooverehead}) despite the \defaultnursurysize nursery size. The GC \%Time column in 
\Cref{tab:macrooverehead} is the percentage of total application time spent in garbage collection -- this value is larger 
than is typical for a mature language/GC stack, however our analysis indicates that this is a function of an unoptimized GC codebase 
and high allocations rates incurred by the baseline implementations for the persistent data-structures. Despite this, the values are still all 
under $12.5\%$ for all benchmarks, indicating that the collector architecture is fundamentally performant. 

These results demonstrate that the constant-factor overheads of the \gc collector, as computed in the theorems/proofs in 
\Cref{sec:theory}, match the results in practice. Although our benchmark applications are limited in size the fundamental properties 
of the collector design, and theoretical guarantees, indicate that these results should hold for larger applications as well, Empirically, 
in \Cref{tab:macrooverall,tab:macrooverehead}, we note that the performance of the \gc collector is largely invariant across the workloads 
and that, even in the face of heavy allocation, the collector does not experience long pauses or is ever out-run by the application.

\subsection{GC Parameter Sensitivity}
\label{sec:parameter-sensitivity}
The \gc collector is designed to provide stable performance characteristics across a range of workloads and allocation patterns.
However, there are fundamental configuration parameters, nursery size and page size, that critically impact the performance of the collector. 

To evaluate the sensitivity of the collector to these parameters we ran a set of experiments varying the nursery size from \mb{1} to \mb{16} 
and page sizes from \kb{2} to \kb{16}. The results of these experiments are shown in \Cref{tab:hpscan}. As our results were consistent across
all benchmarks we show only the measurements for the (largest benchmark) compiler.

\begin{table*}[ht]
\centering
\begin{tabular}{cccccc}
\toprule
Nursery & Page & Total \%GC Time & 50\% (\us{}) & 95\% (\us{}) & 99\% (\us{})\\
\midrule
\defaultnursurysize & \kb{2} & 11.9\% & 209 & 347 & 466 \\
\defaultnursurysize & \kb{4} & 9.9\% & 171 & 211 & 247 \\
\defaultnursurysize & \kb{8} & 9.5\% & 169 & 214 & 239 \\
\defaultnursurysize & \kb{16} & 9.8\% & 186 & 308 & 435 \\
\midrule
\mb{1} & \defaultpagesize & 21.3\% & 77 & 97 & 107 \\
\mb{2} & \defaultpagesize & 16.1\% & 90 & 115 & 125 \\
\mb{4} & \defaultpagesize & 11.8\% & 116 & 146 & 164 \\
\mb{8} & \defaultpagesize & 9.5\% & 169 & 214 & 239 \\
\mb{16} & \defaultpagesize & 10.2\% & 330 & 705 & 1045 \\
\bottomrule
\end{tabular}
\caption{Total GC percentage time and GC pause percentile times on the compiler benchmark (\us{}) for varying nursery sizes (\mb{1}-\mb{16}) and page sizes 
(\kb{2}-\kb{16})}.
\label{tab:hpscan}
\end{table*}
  
As shown in \Cref{tab:hpscan} the nursery size has a significant impact on the collector pause times and total percentage time spent in GC. As we would expect,  
smaller nursery sizes lead to lower pause times as the amount of data copied during each collection is reduced. However, this comes at the cost of increased total 
percentage time spent in GC as the frequency of collections increases. Values of \mb{4} or \mb{8} have similar performance characteristics, \mb{4} has lower pause 
times while \mb{8} has lower total percentage time in GC. Similarly, the sweep over page sizes shows \kb{4} and \kb{8} perform well. Based on these results, we 
use a default nursery size of \defaultnursurysize and a default page size of \defaultpagesize as providing good pause times while keeping total throughput high.

\subsection{Application Performance Distribution}
\label{sec:distribution-analysis}
This section presents a set of controlled experiments that are intended to evaluate the end-to-end statistical behaviors of the \bosque runtime, the 
influence of the \gc collector on this behavior, and, specifically, how close to ideal memoryless execution the \gc runtime comes.

The first experiment examines the end-to-end latency distributions of the core benchmarks (as opposed to simply the collector latency behaviors). 
To evaluate this we created incrementalized versions of the three benchmarks, \cf{nbody-i}, \cf{raytracer-i}, and \cf{db-i}, based on the 
corresponding core benchmarks. These incrementalized benchmarks are designed to run a series of small tasks, each expected with a roughly 
\ms{50} median completion time, mimicking a service responding to various requests. Using these incrementalized benchmarks we can measure the 
end-to-end latency distributions of the applications when run under the \gc collector.

\begin{table*}[ht]
\centering
\begin{tabular}{cccc}
\toprule
Benchmark & 50\% (\ms{}) & 95\% (\ms{}) & 99\% (\ms{})\\
\midrule
nbody-i & 49.2 & 50.0 & 50.4 \\
raytracer-i & 53.4 & 54 & 54.2 \\
db-i & 60.8 & 61.3 & 61.6 \\
\bottomrule
\end{tabular}
\caption{Response percentile times (ms) for the incrementalized versions of the core benchmarks (nbody, raytracer, and db). These 
benchmarks are designed to run a series of small tasks, each expected with a roughly \ms{50} median completion time. The next three 
columns show the $50^{\text{th}}$, $95^{\text{th}}$, and $99^{\text{th}}$ percentile response times for each task on over the benchmarks.}
\label{tab:appdistribution}
\end{table*}

The results of this experiment are shown in \Cref{tab:appdistribution} which shows the $50^{\text{th}}$, $95^{\text{th}}$, and $99^{\text{th}}$
percentile response times for each task on over the benchmarks. As can be seen in this table the response times are tightly clustered, indicating 
consistent performance characteristics. The $50^{\text{th}}$ percentile times for each of the core benchmarks are near \ms{50} as expected while 
the $95^{\text{th}}$ and $99^{\text{th}}$ percentile time tails are very tight, with the $95^{\text{th}}$ percentile being under a $2\%$ 
increase over the $50^{\text{th}}$ percentile times and staying under \ms{51}-\ms{62} (less than $3\%$ slower) even for $99^{\text{th}}$ percentiles. 

This is a remarkable result as it indicates that the \gc collector and runtime provide a very stable and predictable performance profile for the 
applications. Critically, this shows that the stability of the core GC seen in \Cref{tab:macrooverall} translates into well behaved tail-latency 
in the end-to-end application performance as well.

As our idealized runtime model is one where the performance profile for any given task is path-independent, alternatively the runtime behavior is memoryless,
we perform a further experiment to analyze this server application behavior in more depth. Specifically, we want to examine how the distribution of 
response times for each operation changes when:
\begin{enumerate}
    \item Each operation is run in isolation as a uniform workload consisting only of that operation.
    \item Each operation is run as part of a mixed workload consisting of all operations intermixed.
\end{enumerate}

To accomplish this we created a \cf{Server} application that runs tasks from all three core benchmarks in a uniform worklist or in a 
randomly intermixed worklist, while disaggregating the timings into their respective benchmarks, via the following pseudo-code:
\begin{lstlisting}[language=C++]
const List<NbodyTask*> nbodyTasks = ...;
const List<RaytracerTask*> raytracerTasks = ...;
const List<DbTask*> dbTasks = ...;

void recordTiming(ITask* task, GCTimeData* times) {
    if(task->isNbodyTask()) {
        nbodyTimings.record(times);
    } else if(task->isRaytracerTask()) {
        raytracerTimings.record(times);
    } else {
        dbTimings.record(times);
    }
}

void runIsolated() {
    for(size_t i = 0; i < nbodyTasks.size(); i++) { nbodyTasks[i]->run(); }
    for(size_t i = 0; i < raytracerTasks.size(); i++) { raytracerTasks[i]->run(); }
    for(size_t i = 0; i < dbTasks.size(); i++) { dbTasks[i]->run(); }
}

void runMixed() {
    List<ITask*> allTasks = {};
    std::copy(nbodyTasks.begin(), nbodyTasks.end(), std::back_inserter(allTasks));
    std::copy(raytracerTasks.begin(), raytracerTasks.end(), std::back_inserter(allTasks));
    std::copy(dbTasks.begin(), dbTasks.end(), std::back_inserter(allTasks));

    std::random_shuffle(allTasks.begin(), allTasks.end());
    for(size_t i = 0; i < allTasks.size(); i++) { allTasks[i]->run(); }
}   

...

\end{lstlisting}

In this code snippet the \cf{runIsolated} function runs each of the core benchmark tasks in sequence, while the \cf{runMixed} function creates a mixed
worklist of all tasks from all benchmarks and runs them in a random order. In both cases the time taken for each individual task is recorded and then disaggregated 
back into the timings specific to the respective benchmark task kind. The results of this experiment are shown in \Cref{tab:thing}.

\begin{table*}[ht]
\centering
\begin{tabular}{ccccccc}
\toprule
\multicolumn{1}{c}{Benchmark} & \multicolumn{3}{c}{Uniform (\ms{})} & \multicolumn{3}{c}{Mixed (\ms{})} \\
& Average & $1\sigma$ & $2\sigma$ & Average & $1\sigma$ & $2\sigma$\\
\midrule
nbody-i & 50.4 & 0.5 & 1.2 & 51.1 & 0.6 & 1.2 \\
raytracer-i & 53.1 & 0.2 & 0.4 & 53.4 & 0.8 & 1.6 \\
db-i & 52.0 & 0.4 & 0.8 & 52.3 & 0.5 & 1.0 \\
\bottomrule
\end{tabular}
\caption{Analysis of path-independence of operation performance -- \emph{Uniform} columns are times when tasks from a isolated workload runs are measured.
\emph{Mixed} columns are times when tasks from the mixed workload are run and then the timings disaggregated back to individual task kinds.}
\label{tab:thing}
\end{table*}

The Uniform columns in \Cref{tab:thing} show the average and standard deviation times for each operation when run on each benchmark 
workload independently. By construction these workloads are uniform and each task in the workload is expected to take roughly the same 
amount of time. The results in the \emph{Uniform} columns show average response times 
centered around roughly \ms{50} and a $2\sigma$ deviation of \ms{0.4}-\ms{1.2}. The Mixed columns show the same metrics for the operations 
when run in the context of the mixed workload -- that is the task is run alongside and mixed with other tasks but the times 
are disaggregated back into their respective tasks specific categories.

The results in the \emph{Mixed} columns of \Cref{tab:thing} show that in the mixed workload the performance characteristics of individual 
tasks closely resemble the distributions of their uniform counterparts. Specifically the average times are only slightly higher, 
by \ms{0.3} (under $1\%$), and the $2\sigma$ standard deviation is only slightly wider with a range of \ms{1.0}-\ms{1.6}. 
This suggests that the mixed workload does not significantly alter the performance profile of individual tasks, 
supporting the notion of path-independence in our idealized runtime model. 

\subsection{Direct Comparison with SoTA Java GC}
The final experiment is a direct comparison between the \gc collector and modern state-of-the-art low-latency Java garbage collectors on the same benchmark. The binary-trees 
benchmark is a small, but widely used benchmark from the Benchmark Shootout~\cite{bencharkshoot}, that is highly heap intensive. It is designed explicitly to stress GC algorithms 
by creating long-lived data structures while simultaneously allocating at a very high rate. For our purposes it is also possible to implement using exactly same code structure 
in both \bosque and Java allowing for a true 1-1 comparison of Bosque/\gc with an mainstream language and various heavily optimized state-of-the-art GC algorithms.

The data in \Cref{tab:sota} shows the results of this experiment comparing \gc with ZGC~\cite{zgc} and Shenandoah~\cite{shenandoah} (in their default configurations), two 
modern low-latency garbage collectors for Java running with the Java 17 JVM. These collectors are heavily optimized for concurrent and parallel collection, as opposed to 
\gc which is currently a baseline single-threaded implementation which fully pauses the application for the full collection cycle. 

\begin{table*}[ht]
\centering
\begin{tabular}{l|cccccc}
\toprule
Max Heap & \multicolumn{2}{c}{ZGC}& \multicolumn{2}{c}{Shenandoah} & \multicolumn{2}{c}{\gc}\\
& Overhead & Time(\secs{}) & Overhead & Time(\secs{}) & Overhead & Time(\secs{})\\
\midrule
Unlimited & 8.1 & 3.6/4.7 & 12.3 & 2.0/2.9 & 1.3 & 5.7/5.3 \\
1.5$\times$ Live & 3.1 & 7.0/17.3 & 1.5 & 3.4/10.6 & 1.3 & 5.6/5.5 \\
\bottomrule
\end{tabular}
\caption{Direct comparison of \gc with state-of-the-art low-latency Java garbage collectors ZGC and Shenandoah on the binary-trees benchmark. The first column is the max heap size 
allowed, either unlimited or set to $1.5\times$ the live heap size. The next set of columns shown the memory over-provisioning overhead and time taken by each collector
-- ZGC, Shenandoah, and \gc respectively.}
\label{tab:sota}
\end{table*}

The first row in \Cref{tab:sota} shows the results when the max heap size is unlimited allowing the collectors to use as much memory (up to 64 GB) and as many threads as desired 
(with a 16/32 core CPU). The Overhead columns show the memory over-provisioning factor as computed by the max heap size divided by the live heap size. The Time columns show the 
total time taken by the benchmark in terms of wall-clock time and CPU time over all threads (possibly running in parallel). As can be seen, both ZGC and Shenandoah are able to 
complete the benchmark in lower wall-clock time than \gc. However, this performance comes at the cost of significant memory overheads between $8.1\times$ and $12.3\times$ the 
live heap size -- in the case of Shenandoah using \gb{3.3} when compared to \mb{359} for \gc.

The second row shows the result of the benchmark with the max heap size limited to $1.5\times$ the live heap size (\mb{400}). As shown in column 2 of \Cref{tab:sota},
both ZGC and Shenandoah experience severe performance degradation under this configuration with wall-clock time $1.7\times-1.9\times$ higher and CPU time spiking by 
nearly $3.6\times$. Conversely, \gc experiences no performance degradation under this configuration with both wall-clock and CPU times remaining stable. 

Additionally, both ZGC and Shenandoah experience significant issues with GC pauses and application stalls under this configuration. The collectors report degenerate GC runs and 
forced synchronous collections as they are unable to keep up with the allocation rate. Conversely, \gc continues to operate normally with increased pauses due to the higher survival 
rates, and heavy RC workloads, but the fundamental characteristics of the collector prevent the emergence of pathological issues. In fact removing the 
RC decrement phase from the stop-the-world collection, \eg by performing these operations concurrently on a background thread, is sufficient to keep the \gc 
$50^{\text{th}}$ percentile times at \us{119} and even the $99^{\text{th}}$ percentile times under \ms{10}.

These results demonstrate that the \gc collector can provide comparable low-latency performance to state-of-the-art Java garbage collectors while being immune to fundamentally 
pathological behavior tradeoffs that are unavoidable in existing mainstream languages and runtimes.

%% file: related.tex

The fields of memory management and garbage collection are vast topics~\cite{gchandbook}. Thus, in this section we focus on the most relevant 
aspects of these fields as they pertain to the \bosque language and \gc runtime.

\subsection{The Costs of Garbage Collection}
This work is heavily motivated by the analyses in~\cite{distillingcost,understandcost} which explore the costs, including directly incurred and visible 
components as well as second order effects that are diffuse but substantially impact application behavior. Recent work has shifted heavily from optimizing 
for throughput toward the issues of latency and application responsiveness~\cite{rethinkperformance,lxr,zgc,shenandoah,go}. These results enable the use of 
garbage collected languages in application spaces that require low-latency and would have previously had to be written in languages with manual memory 
management. However, as demonstrated empirically and recently theoretically~\cite{pathalogical}, there are inherent trade-offs in the design space of garbage 
collectors that make it impossible to simultaneously optimize for both throughput, starvation, and latency in existing mainstream languages\rvc{.}{ :(}

These results and insights, along with the shift of computing from centralized workloads to service based distributed architectures~\cite{serverlessdef} 
and edge computing have important implications for the design of both garbage collectors and programming languages. As shown in this work, a holistic approach 
to this new paradigm presents unique opportunities for research on garbage collection techniques.

\subsection{Reference Counting Collectors}
The idea of mixing reference counting and tracing collection 
in a single (generational) collector has been explored in previous work~\cite{urc,intrc,immix}. These approaches aim to combine the benefits of both 
techniques, allowing for more efficient memory management.

The integration of generations with reference counting is critical for allowing the \gc collector to avoid touching old objects (aside from the fringe) 
once they have been promoted. Various forms of reference-counting collectors have been explored in the literature recently~\cite{urc,intrc,conservativegc,gcring} and 
and cover many points in the design space. A key issue, as explored in~\cite{conservativegc}, is the treatment of roots as precise or conservative. 
Although, precise roots have great appeal from a collector implementation standpoint, a conservative design presents flexibility and simplification options 
that are practically beneficial in many scenarios. In particular integration into larger software systems, \eg. JavaScript engines~\cite{chakracore,jsc}, 
and enabling aggressive compiler optimization without worrying about maintaining root storage invariants.

\subsection{Multi-Treaded Execution}
The current collector and the \gc runtime are single-threaded. However, there is active work to add structured task-parallel computation to \bosque. In this 
model we, ideally, want to provide fully thread-local allocation and collection. The invariant that the old space is fully reference-counted and only \emph{fringe} 
objects are touched during collection, presents compelling opportunities for parallelism. Fully, thread local allocation is a trivial extension of the current 
model -- simply a per-thread nursery. However, with a structured task-parallel model, it may be possible to model shared inter-thread references via an 
ownership and borrowing system~\cite{lineartypes,singularity} that would allow threads to share references without (or with minimal) synchronization during collection.
In this model each child can perform a thread-local collection, modulo reference count operations, without synchronization. Thus, synchronization between 
threads would be reduced into two critical sections of the GC.

\subsection{Stack and Region Allocation}
The current collector and \gc runtime could be extended to support a region-based memory management model~\cite{dsa,cyclone}. The functional 
nature of \bosque naturally lends itself to simpler region identification and the, already thread-local and page based, structure of the \gc collector make 
the implementation of stack or region allocation more practical than in a language with more complex memory semantics. However, \bosque has another feature, 
a focus on functor libraries~\cite{bosque,linq,javastreams} for collection processing. These libraries provide a single call for applying an operation to 
a \cf{List<T>} or \cf{Map<K, V>}.

Thus, there is also the possibility for a specialized optimizer that understands the semantics of these operations as atomic components instead of a series 
of individual allocations. In particular, with operation like a \cf{map(fn)} which is of type \cf{List<T> -> List<U>} that produces a new collection of the 
same cardinality and all temp values allocated in fn are dead after the call, it is possible precompute the memory needed and reduce all allocations pointer bumps\rvc{.}{!}

\subsection{Ownership GC}
Ownership-based garbage collection is an emerging paradigm that leverages ownership semantics to manage memory more effectively. Notably, the 
Perceus~\cite{perceus,beans} collector uses the type system to detect when an object is no longer used and can be immediately recycled or efficiently 
updated in place. This design choice can produce very efficient executable code from the source, functional, language. However, this is a tradeoff 
in a system like \gc where immediate recycling of objects would break invariants around old/young object locations, such as the impossibility of old$\rightarrow$young 
references, and possibly introduce the need for remembered sets.

\subsection{Tail Latency}
Application latency, and tail-latency in particular, are critical issues in modern computing systems~\cite{go,tailatscale,isogc,usability}.
The garbage collector is a critical component of a runtime system and is often a major source of variance in application performance behavior. Massive 
work has gone into various GC algorithms to reduce their costs -- with a particular focus on latency~\cite{lxr,shenandoah,go,zgc,isogc}. However, in 
a language with mutation, cycles, and semantically observable object identity, there are fundamental limitations to what can be achieved~\cite{pathalogical,distillingcost} 
-- specifically tradeoffs between latency, throughput, and starvation along with the increasing complexity of the memory management implementation.

Conversely the \bosque project, and \gc runtime, present an alternative view where simplicity and simplification of the language semantics open new 
opportunities for garbage collection design. In particular, the results in~\cite{pathalogical} show that it is theoretically impossible for any (mainstream) 
imperative language to simultaneously provide low-latency and high-throughput garbage collection. This places \bosque in a unique position as the only 
language/runtime stack that, by allowing the assumption of very strong invariants about program state, enables the kind of aggressive garbage collection 
design presented in this paper.